\documentclass[
aip,
apl,
superscriptaddress,
amsmath, amssymb,
amssymb,
reprint,
floatfix
]{revtex4-1}

\usepackage{graphicx}
\usepackage{bm}
\usepackage{color}
\usepackage[]{natbib}
\usepackage{amsmath,amssymb,amsfonts}
\usepackage[normalem]{ulem}
\usepackage{hyperref}	

\usepackage{tabularx}	

\begin{document}

\title{Low-temperature tunable radio-frequency resonator for sensitive dispersive readout of nanoelectronic devices}

\author{David J. Ibberson}\thanks{david.ibberson@bristol.ac.uk}
\affiliation{Quantum Engineering Technology Labs, University of Bristol, Tyndall Avenue, Bristol BS8 1FD, UK}
\affiliation{Hitachi Cambridge Laboratory, J.J. Thomson Avenue, Cambridge CB3 0HE, UK}
\author{Lisa A. Ibberson}
\affiliation{Hitachi Cambridge Laboratory, J.J. Thomson Avenue, Cambridge CB3 0HE, UK}
\author{Geoff Smithson}
\affiliation{Cambridge Consultants Ltd., Cambridge Science Park, Milton Road, Cambridge CB4 0DW, UK}
\author{James A. Haigh}
\affiliation{Hitachi Cambridge Laboratory, J.J. Thomson Avenue, Cambridge CB3 0HE, UK}
\author{Sylvain Barraud}
\affiliation{CEA/LETI-MINATEC, CEA-Grenoble, 38000 Grenoble, France}
\author{M. Fernando Gonzalez-Zalba}\thanks{mg507@cam.ac.uk}
\affiliation{Hitachi Cambridge Laboratory, J.J. Thomson Avenue, Cambridge CB3 0HE, UK}

\begin{abstract}
We present a sensitive, tunable radio-frequency resonator designed to detect reactive changes in nanoelectronic devices down to dilution refrigerator temperatures. The resonator incorporates GaAs varicap diodes to allow electrical tuning of the resonant frequency and the coupling to the input line. We find a resonant frequency tuning range of 8.4~MHz at 55~mK that increases to 29~MHz at 1.5~K. To assess the impact on performance of different tuning conditions, we connect a quantum dot in a silicon nanowire field-effect transistor to the resonator, and measure changes in the device capacitance caused by cyclic electron tunneling. At 250 mK, we obtain an equivalent charge sensitivity of 43 $\mu e / \sqrt{\text{Hz}}$ when the resonator and the line are impedance-matched and show that this sensitivity can be further improved to 31 $\mu e / \sqrt{\text{Hz}}$ by re-tuning the resonator. We understand this improvement by using an equivalent circuit model and demonstrate that for maximum sensitivity to capacitance changes, in addition to impedance matching, a high-quality resonator with low parasitic capacitance is desired.
\end{abstract}

\maketitle

High-frequency reflectometry is a technique widely used to study the dynamical properties of nanoelectronic devices due to its enhanced sensitivity and large bandwidth when compared to direct current measurements~\cite{schoelkopf1998radio, Colless2013}. By embedding a device in an electrical resonator, resistive or reactive changes in the device can be inferred from the resonator's response. 

Previous work has shown that, for sensitive detection of resistive changes, good impedance matching between the high frequency line and the resonator, as well as large fractional changes in resistance, are paramount~\cite{Roschier2004a, Muller2013}. Tunable resonators have recently been explored for sensitive capacitance readout~\cite{ares2016sensitive, schupp2018radio}, concluding that optimal sensitivity occurs when the resonator is impedance matched to the line. In this Letter, we extend the work of Ares et al. and demonstrate that for maximal sensitivity to capacitance changes, in addition to an optimally matched resonator, large fractional changes in capacitance and a high internal-Q resonator are essential. We focus on dispersive changes because, for quantum computing technologies, measurement via detection of reactive changes is preferred since it allows for quantum-limited measurements of the qubits \cite{johansson2006fast,tornberg2007dispersive,mallet2009single}.    

To achieve this, we present a tunable high-frequency resonator that remains operational at the base temperature of a dilution refrigerator and allows matching to be achieved at 200~mK. In previous reports, impedance matching was limited to temperatures of 1~K and above\cite{Muller2010, ares2016sensitive}. We couple our resonator to a quantum dot (QD) in a silicon nanowire field-effect transistor (NWFET) and measure changes in the device capacitance due to adiabatic single-electron tunneling events. We observe that sensitive detection of capacitance changes relies on an optimal balance of maximum power transfer to the resonator, maximal internal quality factor and minimized resonator capacitance, and that all three parameters must be considered simultaneously when designing resonators for dispersive readout.

\begin{figure}
\centering
\includegraphics[scale=1]{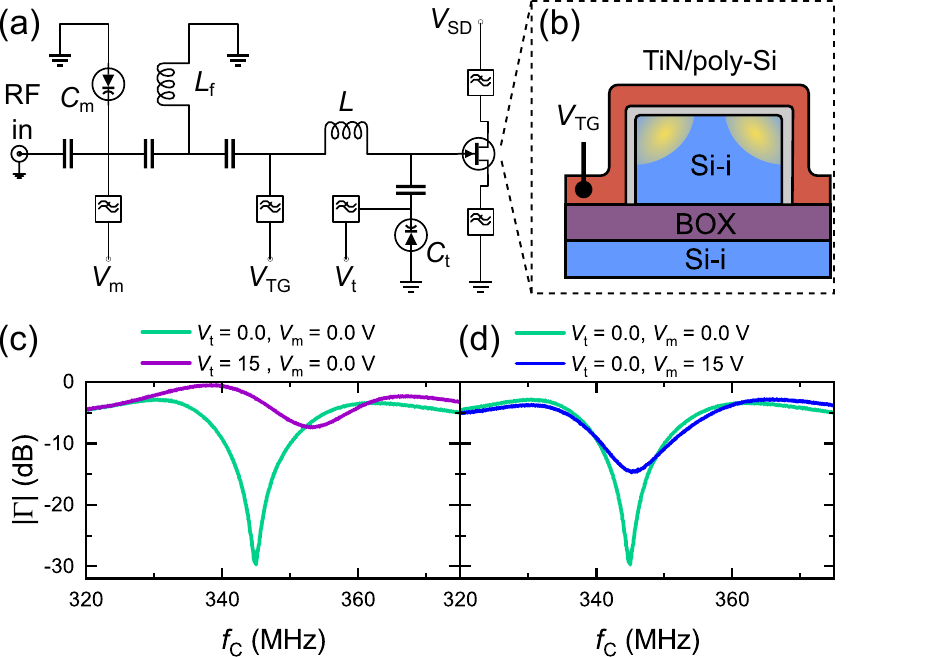}
\caption{Experimental set-up and circuit operation. (a)  Circuit schematic of the tunable resonator. (b) Cross-section schematic of the device perpendicular to the transport direction. Electrostatically-defined QDs form in the corners of the NWFET at low temperatures when $V_\text{TG}$ is close to the threshold voltage~\cite{Voisin2014, chatterjee2018silicon}. (c), (d) Experimental data showing the magnitude of the reflection coefficient as a function of RF carrier frequency at the limits of $V_\text{t}$ and $V_\text{m}$ respectively. Data are acquired with a Vector Network Analyzer (VNA), with the gate of the device connected to the resonator as shown in (a). The DC bias lines for the gate, source and drain are all grounded.}
\label{fig1}
\end{figure}

Our low-temperature resonator is based on an L-match circuit formed by an inductor $L = 270$~nH in series with the parallel combination of the device under study and its parasitic capacitance to ground $C_\text{p}$. We incorporate two reverse-biased GaAs varicap diodes, $C_\text{m}$ and $C_\text{t}$, in a pi-match configuration to provide control over the resonant frequency and matching, see Fig.~\ref{fig1}(a). A shunt inductor $L_\text{f} = 560$~nH provides attenuation of modulating frequencies ($<30$~kHz) to avoid unintentional modulation of the varicap $C_\text{m}$ during sensitivity measurements, see Supplementary Information. The printed circuit board (PCB) is optimized to reduce parasitic capacitances. The resonator is connected to the top gate of a NWFET. This metallic gate wraps around three sides of the nanowire, see Fig.~\ref{fig1}(b), resulting in a large gate-coupling factor, $\alpha=0.95 \pm 0.06$, similar to previously reported values~\cite{Voisin2014,david2018valley}. The results presented in this paper are performed in a cryo-free dilution refrigerator using gate-based reflectometry and homodyne detection \cite{gonzalez2016gate, betz2015dispersively, west2019gate,pakkiam2018single,urdampilleta2018gate}. 

For the typically large impedances of nanoelectronic devices at radio-frequencies, the circuit's resonant frequency is $f_0\approx 1/(2\pi\sqrt{LC_\text{tot}})$ where $C_\text{tot} = C_\text{t}+C_\text{d}+C_\text{p}$. The equivalent impedance at resonance is 

\begin{equation}\label{eqimp}
	Z_\text{eq}\approx\frac{C_\text{tot}^2LR_\text{d}}{{C_\text{m}}^{2}L+C_\text{tot}^3R_\text{d}^2},
\end{equation}

\noindent where $C_\text{d}$ is the state-dependent capacitance of the device and $R_\text{d}$ lumps together the losses in the system including dielectric losses in the device, PCB, and the tuning varicap, see Supplementary Information. The variable capacitance $C_\text{t}$ provides control of $f_0$ and $Z_\text{eq}$ (and therefore the impedance mismatch to the external line) and $C_\text{m}$ provides control of $Z_\text{eq}$ without affecting $f_0$ significantly.  

Fig.~\ref{fig1}(c) demonstrates control of $f_0$ at 55 mK, via the voltage $V_\text{t}$ applied to varicap $C_\text{t}$. We plot the magnitude of the reflection coefficient $\left| \Gamma \right|$ as a function of the carrier frequency $f_\text{c}$ for the limiting varicap bias conditions: $V_\text{t}= 0$ V (high varicap capacitance) and $V_\text{t}= 15$ V (low varicap capacitance). The input RF power $P_c$ is $-85$~dBm. The reflection coefficient is given by $\left| \Gamma \right|=\left|(Z_\text{eq}-Z_0)/(Z_\text{eq}+Z_0)\right|$ where $Z_0$ is the impedance of the external line. The resonant frequency $f_\text{0}$, which corresponds to the frequency at which $\left| \Gamma \right|$ is minimum, increases by 8.35~MHz with the reduction in varicap capacitance. The frequency shift is accompanied by an increase in $\left| \Gamma \right|$ at resonance from -30 to -7~dB, due to the change in impedance mismatch between the device and the external line, see Eq.~\ref{eqimp}.

Fig.~\ref{fig1}(d) shows the effect of $V_\text{m}$ at 55 mK; at resonance $\left| \Gamma \right|$ increases from -30 to -15~dB as $V_\text{m}$ increases from 0 to 15~V, as predicted by Eq.~\ref{eqimp}. A small (0.5~MHz) shift in $f_\text{0}$ accompanies this change. 


\begin{figure}
\centering
\includegraphics[scale=1]{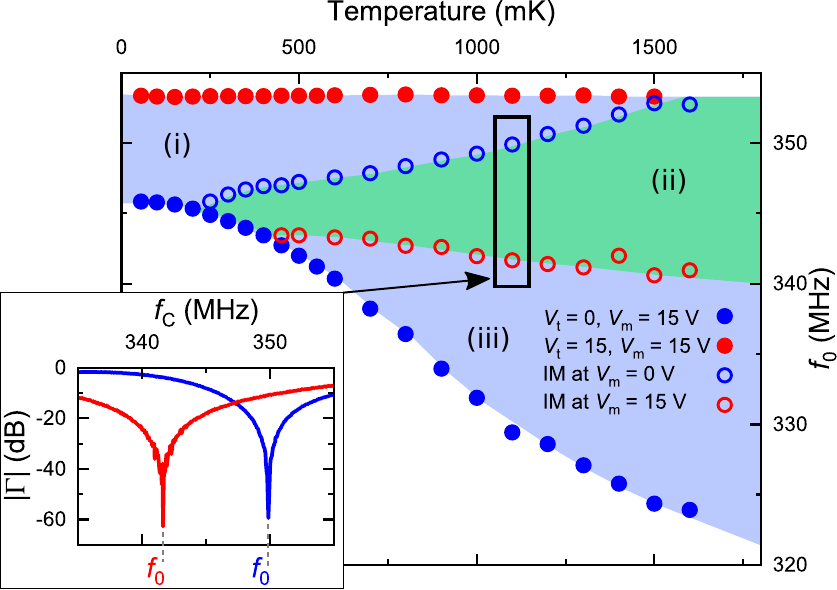}
\caption{Resonant frequency as a function of temperature at the limits of $V_\text{m}$ and $V_\text{t}$. The solid points and the blue shaded region in between indicate the resonant frequency tuning range. Hollow points and the green shaded region (ii) in between represent the frequency range in which impedance matching (IM) is possible. The inset plot shows two of the impedance-matched resonances at 1.1~K, separated in frequency by 8.25 MHz. For all data shown, the source, drain and gate of the device are grounded, with the gate of the device connected to the resonator as shown in Fig.~\ref{fig1}(a).}
\label{fig2}
\end{figure}

We characterize our resonator's performance as a function of temperature. Fig.~\ref{fig2} plots the position of $f_\text{0}$ as a function of temperature for different configurations of the limiting varicap biases. The solid red points plot $f_\text{0}$ at the upper bias limit $V_\text{t} = 15$ V (i.e. minimum $C_\text{t}$ and hence maximum $f_0$). We observe that $f_\text{0}$, and hence $C_\text{t}$, is independent of temperature at this setting. The solid blue points, at which $V_\text{t} = 0$ V (i.e. maximum capacitance), represent the minimum $f_0$. The separation between solid red and blue points, indicated by regions (i-iii), gives the maximum frequency tuning range of the resonator, which increases from 8.4~MHz at 55~mK to 29.0~MHz at 1.5~K. 

Furthermore, the hollow red and blue points track the temperature dependence of $f_\text{0}$ at $V_\text{m} = 15$ and $V_\text{m} = 0$~V, respectively. For each of these points, we adjust $V_\text{t}$ so that the impedance of the resonator matches the impedance of the line. At any frequency between these hollow points, in the green-shaded region (ii), impedance matching can be achieved by changing $V_\text{t}$ and $V_\text{m}$. We achieve impedance matching down to 200 mK, see Supplementary Information. The frequency range for matching increases with temperature up to 12.3~MHz at 1.5~K. 

\begin{figure}
\centering
\includegraphics[scale=1]{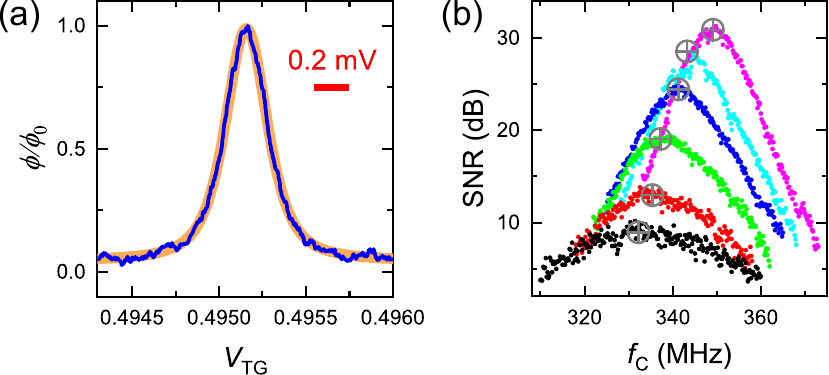}
\caption{Coulomb oscillation and sensitivity optimization. (a) The dot-to-reservoir transition used to measure sensitivity, observed in the normalised phase response of the resonator $\phi/\phi_0$, where $\phi_0$ is the maximum phase value recorded. The data is plotted in blue and our fit in orange, drawn with a thicker line for visual clarity. The input RF power is $-95$~dBm, temperature is $50$~mK and $V_\text{SD}=-10$ mV. The red line indicates the peak-to-peak amplitude of the top-gate modulation signal. (b) Optimization of the RF carrier frequency for different $V_\text{t}$ varicap biases. Black, red, green, blue, cyan and pink correspond to $V_\text{t}=0,~1.5,~ 3,~4.6,~6.5,~15$~V respectively. $V_\text{m}=0$ for all, at a temperature of 1.1~K, with input RF power $-90$~dBm and a spectrum analyzer resolution bandwidth of 2~Hz. The maximum SNR is marked for each data set with a grey circle.}
\label{fig3}
\end{figure}

To benchmark the resonator, we follow the standard procedure for measuring charge sensitivity\cite{schoelkopf1998radio, gonzalez2015probing, ahmed2018radio, Roschier2004a}. We bias the top gate electrode $V_\text{TG}$ to the single-electron charge transition shown in Fig.~\ref{fig3}(a) and apply a small-amplitude sinusoidal modulation (indicated by the red bar in Fig.~\ref{fig3}(a)) to ensure a linear response. The oscillatory change in device capacitance produces amplitude modulation of the carrier and results in side-bands in the frequency domain. Sensitivity to capacitance changes is inversely proportional to the signal-to-noise ratio (SNR) of the side-bands \cite{ahmed2018radio, schoelkopf1998radio, Roschier2004a, Muller2013}. We choose this transition because it is purely capacitive and hence produces a purely dispersive shift of the resonant frequency. To demonstrate this, we fit a linear combination of an inverse square cosh function, due to the tunneling capacitance~\cite{Mizuta2017}, and a Lorentzian, due to tunnel-rate broadening~\cite{Cottet2011, House2015}. We find that the peak is predominantly Lorentzian with a tunnel rate of $\gamma=40\pm3~$GHz. Since $\gamma>>f_\text{c}$, electrons tunnel in and out of the QD adiabatically, producing a purely capacitive response~\cite{ahmed2018thermo}. This result is important because non-adiabatic transitions, which occur when $\gamma$ $\approx$ $f_\text{c}$, produce dissipative effects that can mask the measurement of sensitivity to capacitance changes~\cite{gonzalez2015probing, Esterli2018}. Power broadening of the transition starts to dominate the width when the carrier power $P_\text{c} > -93$~dBm and hence in this figure we use $P_\text{c} = -95$~dBm. 

We optimize $V_\text{TG}$ and $P_\text{c}$ to maximize the SNR and improve the sensitivity. We find the optimal $V_\text{TG}$ at the points of maximum $\left|d\phi/dV_\text{TG}\right|$ and the optimal power at $P_\text{c}=-90$~dBm. The dependence of the SNR of the sidebands on $f_\text{c}$ is explored in Fig.~\ref{fig3}(b). The maximum SNR is expected at the natural frequency of oscillation of the resonator. Since $f_0$ varies with $V_\text{t}$, we measure the SNR versus $f_\text{c}$ curves for multiple $V_\text{t}$ values from 0~V (black dots) to 15~V (pink dots).  As we increase $V_\text{t}$, and hence decrease $C_\text{t}$, the SNR increases and shifts to higher frequencies. To obtain the optimum $f_\text{c}$ values, we extract the maxima of these curves by fitting an asymmetric Lorentzian to each data set. The maxima for each optimized configuration are indicated by grey circles in Fig.~\ref{fig3}(b). 


\begin{figure}
\centering
\includegraphics[scale=1]{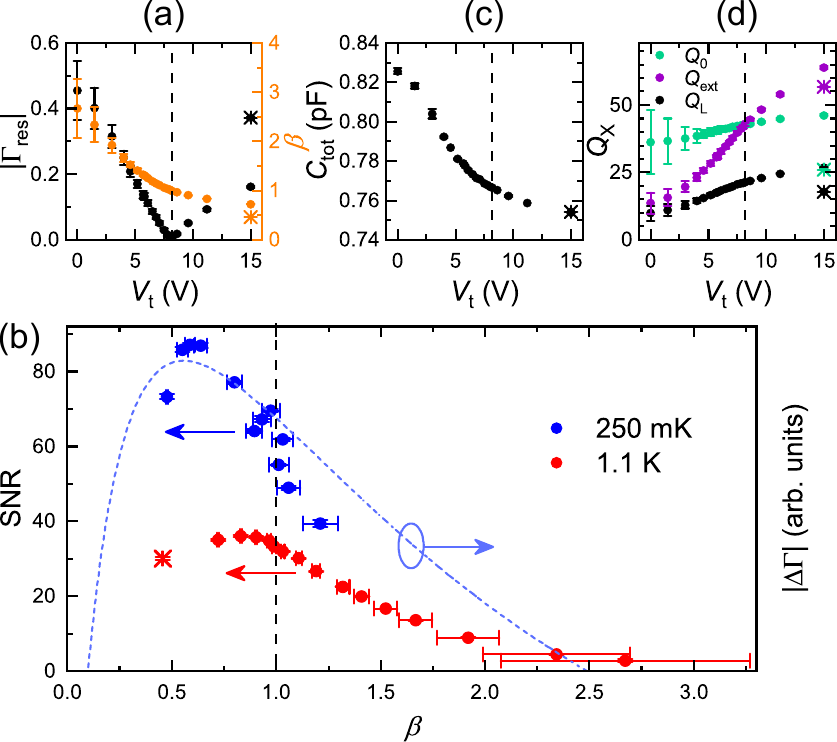}
\caption{Determining the optimum impedance matching conditions for sensitivity measurements. (a) Change in the impedance mismatch as $V_\text{t}$ is varied, characterized by $\left| \Gamma_\text{res} \right|$ (black dots) and the coupling coefficient $\beta$ (orange dots). $V_\text{m}=0$ V for all points except one starred point in each plot, where $V_\text{m}$ was changed to 15~V (in the 1.1~K data only). (b) SNR plotted against the coupling coefficient at 250~mK (blue) and 1.1~K (red). The dashed curve plots $\left| \Delta \Gamma \right|$, obtained using a simplified circuit model. (c) Change in the total capacitance of the resonator-device system ($C_\text{tot}$) as $V_\text{t}$ is varied. (d) Change in the internal, external and loaded quality factors of the resonator as $V_\text{t}$ is varied. Note that (a), (c), and (d) plot measurements at 1.1~K.}
\label{fig4}
\end{figure}

We now investigate the effects of impedance mismatch on sensitivity to capacitance changes by varying the bias of the varicaps. Primarily, we vary $V_\text{t}$ since it provides greater tunability at low temperature. A measure of the impedance mismatch is provided by the coupling coefficient $\beta = Q_\text{0}/Q_\text{ext}$, where $Q_\text{0}$ and $Q_\text{ext}$ are the internal- and external-Q values respectively. The resonator is impedance matched to the line at $\beta=1$. For $\beta<1$, the power is predominantly dissipated in the resonator and the system is undercoupled, whereas for $\beta>1$, the power is predominantly dissipated in the external line and the system is overcoupled. We calculate $\beta$ from the reflection coefficient measured at the resonant frequency, which we denote $\left| \Gamma_{\text{res}} \right|$ \cite{pozarmicrowavebook}.

Fig.~\ref{fig4}(a) plots $\beta$ together with $\left| \Gamma_\text{res} \right|$ as a function of $V_\text{t}$, for $V_\text{m}=0$~V at 1.1~K. At $V_\text{t}=8.2$~V the resonator and the line are impedance matched, $\left| \Gamma_{\text{res}} \right|$ is at a minimum and $\beta=1$.  

We plot the maximum voltage-SNR as a function of $\beta$ at 250~mK and 1.1~K in Fig.~\ref{fig4}(b). Both data series present similar trends, but due to temperature broadening of the dot-to-reservoir transition at 1.1~K \cite{ahmed2018thermo}, the 250~mK SNR values are higher. The voltage of the RF signal at the device gate, the voltage drop, is directly proportional to $Q_\text{L}$ and, at 1.1 K, increases from 140 to 379~$\mu$V as $\beta$ decreases over our experimental range (and from 193 to 334~$\mu$V at 250~mK), see Supplementary Information. It is clear for both temperatures that the maximum SNR does not occur at the point of impedance matching. Instead, in this particular experiment, it occurs in the undercoupled regime. For comparison, at 250 mK, we obtain a charge sensitivity of $43.0 \pm 0.4~ \mu e / \sqrt{\text{Hz}}$ at $\beta=1$, whereas the best sensitivity is $30.5 \pm 0.2~ \mu e / \sqrt{\text{Hz}}$. At 1.1~K, the sensitivity at $\beta=1$ is $80.0 \pm 0.4~ \mu e / \sqrt{\text{Hz}}$ and the best sensitivity is $73.8 \pm 0.5~ \mu e / \sqrt{\text{Hz}}$. This is different to what has been reported previously for dispersive changes~\cite{ares2016sensitive} where maximum sensitivity was achieved at the point of impedance matching. Our data shows that, for capacitive sensing, this is not in general the case and additional factors beyond matching need to be considered.

To gain insight into the conditions that govern sensitive capacitive sensing, we use an equivalent circuit model for our resonator, see Supplementary Information. We calculate the effect of a change in device capacitance $\Delta C_\text{d}$ on $\Gamma$ as $\left|\Delta\Gamma\right|= \left|\Delta C_\text{d}\cdot\partial\Gamma/\partial C_\text{d}\right|$. The measured SNR is directly proportional to $\left|\Delta\Gamma\right|$, given by $\text{SNR}=|\Delta\Gamma|\times \sqrt{P_\text{c}/P_\text{N}}$, where $P_\text{N}$ is the noise power. This approach provides a method to assess the most important parameters for dispersive readout. In the case $R_\text{d}^2 \gg C_\text{m}^2L/C_\text{tot}^3$, the change in reflection coefficient reads

\begin{equation}\label{theory}
\left| \Delta  \Gamma \right| \approx \frac{2 Z_\text{eq} Z_0}{(Z_\text{eq}+Z_0)^2} \times \frac{\Delta C_\text{d}}{C_\text{tot}} \times Q_0.
\end{equation}

For our circuit configuration, we see that $\left| \Delta  \Gamma \right|$ is dependent on three factors. Firstly, on the impedance mismatch between the resonator and the line. Secondly, on the relative capacitance change of the device to the total capacitance of the resonator, and thirdly, on the internal quality factor of the resonator $Q_0$. $\left| \Delta  \Gamma \right|$ can also be improved quadratically by increasing the gate-coupling factor $\alpha$ of the device \cite{gonzalez2015probing, Mizuta2017}. Note that the quality factor does not affect the sensitivity to resistive changes\cite{Muller2013}. As we describe below, our control parameter $V_\text{t}$ affects $Q_0$ and $C_\text{tot}$ as well as the matching, and hence the maximum sensitivity can occur away from $\beta=1$.

In Fig.~\ref{fig4}(c) and (d) these dependencies are demonstrated experimentally. We calculate how $C_\text{tot}$ varies with $V_\text{t}$ by measuring $f_0$ from the experimental data and using $f_0\approx 1/(2\pi\sqrt{LC_\text{tot}})$. As we increase $V_\text{t}$, and therefore decrease $\beta$, the varicap capacitance $C_\text{t}$ decreases and hence $C_\text{tot}$ decreases, see Fig.~\ref{fig4}(c). Next, we study the effect of $V_\text{t}$ on $Q_0$. To do so, we first extract the loaded quality factor $Q_\text{L}$ from the ratio between the resonant frequency and the full width at half maximum of the reflection coefficient against frequency, $Q_\text{L}=f_0/\Delta f$. We then calculate the internal as well as the external quality factors, $Q_0=(1+\beta)Q_\text{L}$ and $Q_\text{ext}=Q_0/\beta$ respectively, and plot them as a function of $V_\text{t}$ in Fig.~\ref{fig4}(d). We observe that as we increase $V_\text{t}$, and therefore decrease the coupling $\beta$, both the internal and external quality factors increase. Hence as we reduce $\beta$ below 1, although the $(2 Z_\text{eq}Z_0)/(Z_\text{eq}+Z_0)^2$ term decreases, $C_\text{tot}$ decreases and $Q_0$ increases, which lead to an increased SNR even though the impedances are mismatched. We note that $Q_0$ depends on the varicap capacitance and hence varies with $V_\text{t}$.


To further support our experimental evidence, in Fig.~\ref{fig4}(b) we compare the measured SNR at 250~mK with a calculation of $\left|\Delta\Gamma\right|$, see dashed line. Although Eq.~\ref{theory} shows a maximum at $\beta=1/3$, a full numerical calculation using $C_\text{m} = 10$~pF and $R_\text{d} = 30$~k$\Omega$ reproduces the 250~mK data well, showing a maximum near $\beta \approx 0.6$, see Supplementary Information for details of the calculation. The position of the maximum at 1.1~K cannot be reproduced with realistic device parameters and including additional parasitic elements in the circuit model may be necessary.

Overall, in order to maximize $\left|\Delta\Gamma\right|$, it is necessary to match the resonator's impedance to the line, but this should not be achieved by reducing the internal quality factor of the resonator, or increasing the capacitance. It is essential to optimize these three terms simultaneously when designing optimal resonators for dispersive sensing.


In conclusion, we have demonstrated a low-temperature tunable RF resonator based on GaAs variable capacitors. Whereas previous reports have been limited to temperatures above 1~K~\cite{ares2016sensitive,Muller2010}, our optimized design enables resonant frequency tuning down to 55~mK and can be impedance matched down to 200~mK. The method can be generally applied to the dispersive measurement of nanoelectronic devices with parametrically variable reactance~\cite{Johansson2006,Shevchenko2008,west2019gate,pakkiam2018single,urdampilleta2018gate} and may be suitable for high-sensitivity frequency multiplexing schemes~\cite{hornibrook2014frequency}. 

Furthermore, we have explored how capacitive sensing depends on coupling
conditions and highlighted three main features an electrical resonator should possess for sensitive capacitive readout: good matching to the line, high fractional capacitive change and a high internal Q. If these elements are changed simultaneously, maximum sensitivity will not necessarily occur at the point of impedance matching.

See Supplementary Information for a detailed description of the resonator circuit, the theoretical model and the experimental technique for impedance matching.


This research has received funding from the European Union's Horizon 2020 Research and Innovation Programme under grant agreement No 688539 (http://mos-quito.eu) and the Winton Programme of the Physics of Sustainability. DJI is supported by the Bristol Quantum Engineering Centre for Doctoral Training, EPSRC grant EP/L015730/1. We thank N. J. Lambert for useful discussions.

\bibliography{Tunable}

\clearpage
\part*{Supplementary Information}

\renewcommand{\arraystretch}{1.2} 
\setcounter{figure}{0}
\setcounter{equation}{0}
 
\renewcommand{\thesection}{S\arabic{section}}   
\renewcommand{\thetable}{S\Roman{table}}   
\renewcommand{\thefigure}{S\arabic{figure}}
\renewcommand{\theequation}{S\arabic{equation}}

\section{Circuit Schematic}\label{Sup_CircuitSch}
\begin{figure}[b]
\centering
\includegraphics[scale=1]{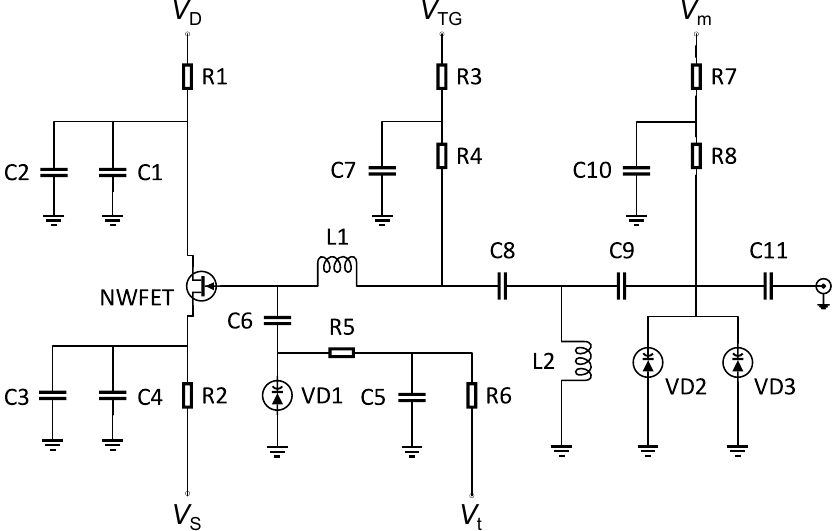}
\caption{Full schematic of the circuit. `NWFET' refers to the silicon nanowire field effect transistor measured.}
\label{Sup_Circuit}
\end{figure}

The full circuit schematic of the resonator used in the measurements reported in the main paper is shown in Fig.~\ref{Sup_Circuit}. The component specifications are detailed in Table \ref{Components}. 

\begin{table}
\caption{Specification of the components used in the tunable matching circuit.}
\newcolumntype{L}{>{\raggedright\arraybackslash}X}
\begin{tabularx}{\columnwidth}{X L}
\hline \hline 
\multicolumn{1}{c}{\textbf{Label in Fig.~\ref{Sup_Circuit}}} & \multicolumn{1}{c}{\textbf{Specification}}\\
\hline

R3 & 1~k$\Omega$ 0603 thin film resistor, TE-Connectivity RP73D1J1K0BTDG \\

R1, R2, R4, R8 & 10~k$\Omega$ 0603 thin film resistor, TE-Connectivity RP73D1J10KBTDG \\

R5, R6, R7 & 100~k$\Omega$ 0603 thin film resistor, TE-Connectivity RP73D1J100KBTDG \\

C6	& 1~pF 0603 NPO COG capacitor, KEMET CBR06C109BAGAC \\

C1, C4, C8, C9, C11	& 100~pF 0603 NPO COG capacitor, Murata GRM1885C1H101JA01 \\

C7	& 1~nF 0603 NPO COG capacitor, Murata GRM1885C1H102JA01 \\

C2, C3, C5, C10	& 10~nF 0603 NPO COG capacitor, KEMET C0603C103J3GACTU \\

L1	& 270~nH $\pm 5$\% 0805 TDK B82498F3271J001 inductor \\

L2	& 560~nH $\pm 5$\% 0805 TDK B82498F3561J001 inductor \\

VD1	& MA46H200 MACOM varicap diode \\

VD2, VD3 & MA46H204 MACOM varicap diode \\
\hline \hline
\end{tabularx}
\label{Components}
\end{table}

The circuit is required to perform three functions: match the NWFET impedance to the dilution refrigerator RF system, enable biases to be applied to the NWFET, and enable voltage control of the varicap diodes. The varicap diodes allow electronic tuning of both the resonant frequency of the circuit, and the impedance mismatch between the circuit and the refrigerator RF system.

The source and drain bias for the NWFET are applied via two DC lines containing a combination of series resistors (R1 and R2) and shunt capacitors (C1, C2 and C3, C4) that provide low-pass filtering. These attenuate noise pickup and provide static protection for the device. The RC time constant of both lines is 1~ms. The capacitors C1 and C4 ensure that we have an RF short circuit on the source and drain of the NWFET. The bias for the top-gate of the NWFET is applied through a much smaller RC time constant of 1~$\mu$s (R3, C7) which enables modulation of the top-gate voltage up to $\approx$ 30 kHz without significant attenuation. Resistor R4 is included to prevent the bias line and associated filtering from loading the RF signal. 

Additional DC lines provide the varicap biases through series resistors (R6 and R7) and shunt capacitors (C5 and C10) so that each line has an RC time constant of 1~ms to attenuate noise pickup. Additional series resistors are used to isolate the RF circuitry and associated filtering from the varicap bias lines. Resistor R8 provides isolation for VD2 and VD3 and R5 provides isolation for VD1. The isolating resistors (R4, R5, R8) must be sufficiently large to minimize loading of the RF circuitry (which would result in the attenuation of the RF signal and hence reduced sensitivity) but small enough to minimize the voltage drop across them due to gate current or varicap reverse leakage current, and to minimize Johnson noise.

In addition to the requirements outlined above, it is also necessary to ensure that there is sufficient isolation between the top-gate modulation waveform, applied during the sensitivity measurements discussed in the main paper, and the varicaps. This isolation is important to ensure that the sidebands measured in the frequency domain result from single electron tunneling events rather than from modulation of the varicap biases. Also, the isolation circuitry must not attenuate the top-gate modulation signal amplitude. Isolation between the top-gate modulating signal and the two parallel varicaps is provided by the reactive potential divider C8 and L2, which has low loss at RF frequencies and high attenuation at the top-gate modulating frequency ${\sim}100$~Hz. The high impedance of C8 at the frequency used to modulate the top-gate ensures that the shunt inductor L2 does not attenuate the top-gate waveform amplitude. We select the value of L2 to be self-resonant at the RF frequency to ensure negligible loss of RF signal. Similarly, isolation between the top-gate modulating signal and the single varicap diode is provided by a reactive potential divider formed by C6, R5, C5, R6. The high impedance of C6 at the top-gate modulating frequency prevents the attenuation of the top-gate waveform. 

The circuit is fabricated on low-loss Rogers RO4003C material which has a low dielectric constant ($\epsilon_r=3.55$, $\tan\delta=0.0021$). The transmission lines are implemented in microstrip. The RF system of the dilution refrigerator has a characteristic impedance of 50 $\Omega$ and the circuitry on the low impedance side of the inductor L1 is implemented with 50 $\Omega$ tracks. The high impedance side of the inductor L1 has a characteristic impedance that is significantly greater than 50 $\Omega$, so the tracking here is focused on the minimization of parasitic capacitance by using the minimum track width possible.

Thin film resistors and NPO COG capacitors have been found to perform well at low temperatures. The two inductors were required to be physically small in order to be compatible with a range of dilution refrigerators. High-Q 0805 chip inductors are a compromise between performance and size; potentially a sensitivity improvement could be achieved by using a larger air core inductor in location L1.

\section{Simplified Model of the Resonator Circuit - Derivation of sensitivity}\label{Sup_Model}

\begin{figure}
\begin{center}
\includegraphics[scale=1]{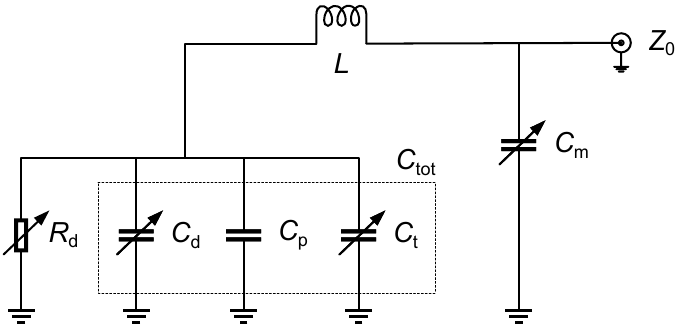}
\end{center}
\caption{Schematic of the simplified circuit used to model the resonator and device. Note $Z_0$ denotes the impedance of the external line.}
\label{simpleModel}
\end{figure}

The circuit model used to obtain Eqs.~1 and 2, and the fit in Fig.~4(b) in the main paper, is shown in Fig.~\ref{simpleModel}. At the frequencies used in the manuscript, the complexity of the circuit can be reduced down to the pi-match design formed by the inductor, $L$, and the variable capacitances $C_\text{tot}$ and $C_\text{m}$. We model the device by a variable capacitor $C_\text{d}$ in parallel with a resistance. $C_\text{d}$ corresponds to the tunneling capacitance~\cite{Esterli2018}. $R_\text{d}$ represents the dielectric losses in the device gate oxide and buried oxide, plus the dielectric losses in the PCB and the tuning varicap $C_\text{t}$. Finally, $C_\text{p}$ is the combined parasitic capacitance of the device and PCB. The three parallel capacitances can be lumped together as $C_\text{tot}=C_\text{t}+C_\text{p}+C_\text{d}$. We name the total impedance of the circuit $Z$. 


We calculate the resonant frequency of the circuit and find that for $R_\text{d}^2\gg aL/C_\text{tot}$, where $a = C_\text{m}/C_\text{tot}$, the system resonates at

\begin{equation}
	\omega_0\approx\sqrt{\frac{1}{LC_\text{tot}}}.
\end{equation}

The equivalent impedance at the resonance, $Z_\text{eq}=Z(\omega_0)$, is
\begin{equation}\label{Z1}
	Z_\text{eq} =\frac{C_\text{tot}^2LR_\text{d}}{C_\text{m}^2L-2C_\text{m}C_\text{tot}L+C_\text{tot}^2(L+C_\text{tot}R_\text{d}^2)},
\end{equation}

\noindent which, in the case $C_\text{m} \gg C_\text{tot}$, simplifies to $\frac{C_\text{tot}^2LR_\text{d}}{C_\text{m}^2L+C_\text{tot}^3R_\text{d}^2}$.

Regarding sensitivity to capacitance changes, we consider the figure of merit $\left| \Delta\Gamma \right|$ which corresponds to the absolute change in the reflection coefficient $\Gamma$ with a given change in device capacitance $\Delta C_\text{d}$,

\begin{equation}\label{dGamma}
	\left| \Delta\Gamma \right| =\left|\frac{\partial\Gamma}{\partial C_\text{d}}\Delta C_\text{d}\right|.
\end{equation}

The signal-to-noise voltage ratio (SNR) is directly proportional to this figure of merit

\begin{equation}\label{SNReqn}
	\mathrm{SNR}=\left| \Delta\Gamma \right| \times\sqrt{P_\text{c}/P_\text{N}},
\end{equation}

\noindent where $P_\text{c}$ corresponds to the power applied to the resonator and $P_\text{N}$ to the noise power of the measurement. We calculate $\left| \Delta\Gamma \right|$ for the case $R_\text{d}^2\gg a^2L/C_\text{tot}$ and obtain

\begin{equation}\label{sense}
	\left| \Delta\Gamma \right| \approx \frac{2Z_\text{eq}Z_0}{(Z_\text{eq}+Z_0)^2}R_\text{d}\sqrt{\frac{C_\text{tot}}{L}}\frac{\Delta C_\text{d}}{C_\text{tot}}.
\end{equation}

Here, $R_\text{d}\sqrt{C_\text{tot}/L}$ corresponds to the internal quality factor of the resonator, $Q_0$. Eq.~\ref{sense} can be conveniently expressed in terms of the coupling coefficient $\beta \approx Z_0R_\text{d}C_\text{tot}/L$ as

\begin{equation}\label{sense2}
	\left| \Delta\Gamma \right| \approx 2\Delta C_\text{d}\frac{\sqrt{\beta}}{(1+\beta)^2}\frac{R_\text{d}}{L}\sqrt{R_\text{d}Z_0}.
\end{equation}

\begin{figure}
    \centering
    \includegraphics{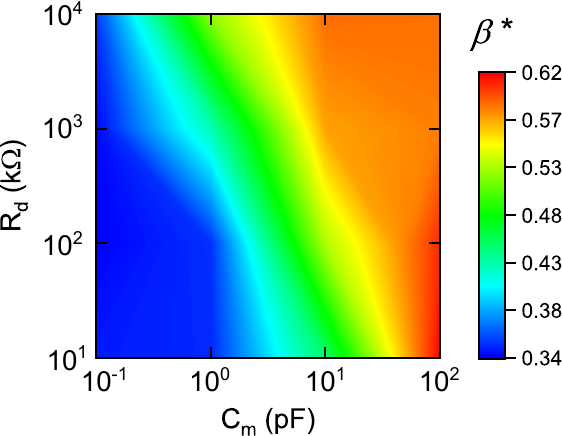}
    \caption{Position in $\beta$ of the maximum $\left| \Delta\Gamma \right|$, $\beta^*$, as the circuit losses $R_\text{d}$ and matching varicap capacitance $C_\text{m}$ are varied. We observe that as either $R_\text{d}$ or $C_\text{m}$ are increased, $\beta^*$ also increases. Given a value for $\beta^*$, there are a range of $R_\text{d}$ and $C_\text{m}$ combinations that can produce this value. Data are calculated by performing a numerical analysis of the simplified circuit model shown in Fig.~\ref{Sup_Model}.}
    \label{betamax}
\end{figure}

Eq.~\ref{sense2} presents qualitatively all the general features of our data in Fig.~4(b). It shows a maximum at a particular $\beta$, in this case at $\beta=1/3$. The shifting position in $\beta$ of the maximum, observed between the two temperatures in Fig.~4(b), can be reproduced by numerically solving Eq.~\ref{dGamma}. We show the results of this in Fig.~\ref{betamax}, where we plot the position in $\beta$ of maximum $\left| \Delta\Gamma \right|$, $\beta^*$, as a function of $R_\text{d}$ and $C_\text{m}$. At low $R_\text{d}$ and $C_\text{m}$, the numerical solution reproduces the analytical solution and as both parameters increase, $\beta^*$ shifts to higher values. We use $R_\text{d}=30$~k$\Omega$ and $C_\text{m}=10$~pF to fit the 250~mK data in Fig.~4(b) which presents a maximum at $\beta^*=0.59$. The position of the maximum at 1.1~K cannot be reproduced with realistic device parameters. Additional circuit elements in the equivalent circuit of Fig.~\ref{Sup_Model} may be necessary. 

\section{Voltage Drop at the device gate}

 In this section we explain the dependence of $\left| \Delta\Gamma \right|$ on the voltage drop at the device gate, $v_\text{drop}$. For a total applied power, $P_\text{c}={v_{0}}^2/2Z_0$ where $v_0$ is the RF voltage at the input of the resonator, a fraction of the power,  $K=1-\left|\Gamma\right|^2$, is dissipated in $R_\text{d}$, such that

\begin{equation}
	K\frac{v_0^2}{2Z_0}=\frac{v_\text{drop}^2}{2R_\text{d}}.
\end{equation}

Hence, the voltage drop at the device is directly proportional to the applied voltage, given by

\begin{equation}\label{sensevdrop}
	v_\text{drop}=2Q_\text{L}v_0
\end{equation}

\noindent where $Q_\text{L}$ is the loaded quality factor of the resonator. Now, we can rewrite Eq.~\ref{sense} as

\begin{equation}\label{sense_volt}
		\left| \Delta\Gamma \right| \approx \frac{1}{2}\left(\frac{v_\text{drop}}{v_0}\right)^2\omega_0Z_0\Delta C_\text{d}.
\end{equation}

From Eq.~\ref{sense_volt} we understand that, for optimal sensitivity, a resonator that maximises the voltage drop at the gate for a given input voltage is desired; in other words, a well-matched, high-$Q_0$ resonator. The dependence of $\left|\Delta\Gamma \right|$, and hence SNR, on $Q_\text{L}$ is quadratic. Note that, although increasing the input voltage will increase the SNR linearly (Eq.~\ref{SNReqn}), at least until saturation, this approach will not increase $\left| \Delta\Gamma \right|$ and therefore is a much less effective way of improving the readout SNR than increasing $Q_\text{L}$.

\begin{figure}
    \centering
    \includegraphics{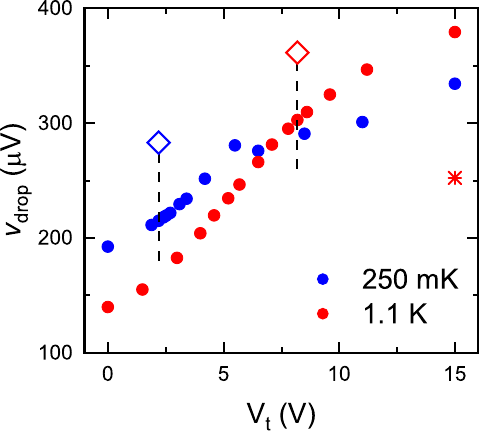}
    \caption{Voltage drop at the device as $V_\text{t}$ is varied for 250~mK and 1.1~K. The dashed vertical lines intersecting each set of points indicate the bias voltage corresponding to impedance matching. $V_\text{m}$ = 0~V for all data points except one starred point at 1.1~K for which $V_\text{m}$ = 15~V.}
    \label{voltagedrop}
\end{figure}

Finally, in Fig.~\ref{voltagedrop}, we plot the voltage drop as a function of $V_\text{t}$ using Eq.~\ref{sensevdrop} and the $Q_\text{L}$ data presented in Fig.~4(d). We calculate $v_0$ from the applied power $P_\text{c}=-93$~dBm. We observe that the voltage drop increases as $V_\text{t}$ is increased.

\section{Experimental Impedance Matching}\label{phasedata}

\begin{figure}
\begin{center}
\includegraphics[scale=1]{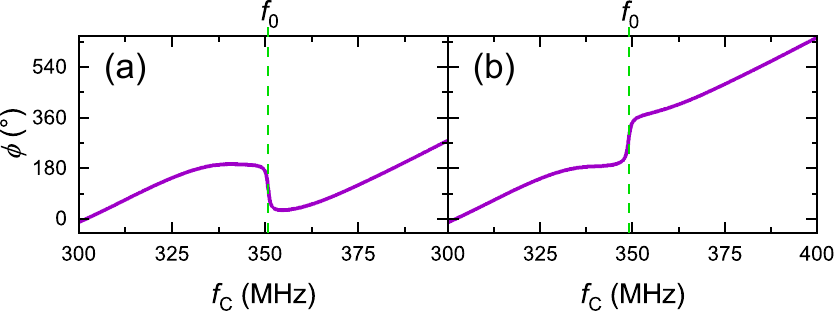}
\end{center}
\caption{Phase of the reflection coefficient as a function of carrier frequency, (a) in the undercoupled regime, (b) in the overcoupled regime. In (a), $V_\text{t}=9.6$ V giving a resonant frequency $f_0=350.82$ MHz. In (b), $V_\text{t}=7.1$ V giving a resonant frequency $f_0=348.94$ MHz. Both plots are obtained with $V_\text{m}=0$ V, an RF input power of -85 dBm and at a temperature of 1.1~K.}
\label{phaseplots}
\end{figure}

We tune the varicap voltages to near-perfect impedance matching by using the Vector Network Analyzer (VNA) to analyze the complex reflection coefficient $ \Gamma=\left| \Gamma \right|e^{i\phi}$ of the reflected signal as a function of frequency, $f_\text{c}$. The magnitude $\left| \Gamma \right|$ (see for example Fig.~1(c, d) in the main paper) is minimized when the impedances match, so firstly, we adjust the varicap biases roughly to find this minimum. Observing the phase of the signal, $ \phi$, shows this point more precisely, allowing fine-tuning of the varicap biases. This is due to the sudden change in the gradient of $\phi$ against $f_\text{c}$, at the resonant frequency $f_0$, when changing between the under- and over-coupled regimes, as illustrated by Fig.~\ref{phaseplots}. This gradient is negative in the undercoupled regime, as observed in (a), and positive in the overcoupled regime, as observed in (b). At near-perfect matching, the phase becomes noisy at $f_0$, appearing to switch rapidly between the two regimes, because the magnitude of the reflected signal is minimal. This technique was performed without prior calibration of the VNA to the line and resonator, which is evidenced by the accumulation of phase with increasing frequency observed in these plots. We varied $V_\text{t}$ and $V_\text{m}$ in steps of 0.1 V, which sets the precision of impedance matching in this experiment.

\section{Impedance matching and tunability at 200~\lowercase{m}K}

\begin{figure}
    \centering
    \includegraphics{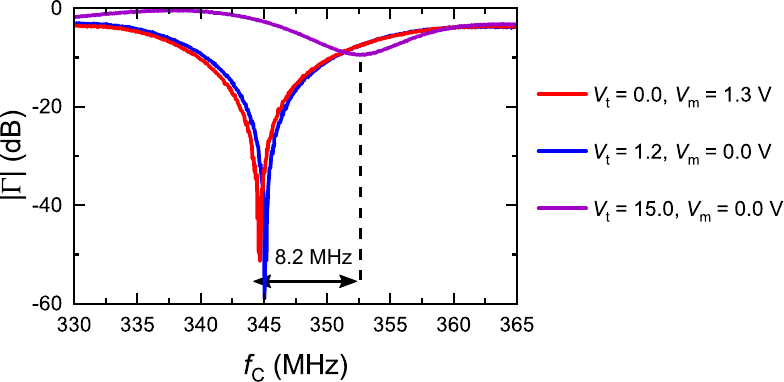}
    \caption{Experimental data showing two impedance matched resonances (red and blue) at 200~mK. The purple trace gives the resonance at the upper limit of frequency tunability.}
    \label{perfectresonances200}
\end{figure}

The lowest temperature at which impedance matching is observed is 200~mK. Fig.~\ref{perfectresonances200} presents the resonances for two varicap bias configurations that produce impedance matching at 200~mK (red and blue lines). We also include the resonance at the upper limit of $V_\text{t}$ (purple) to illustrate the maximum frequency tunability of 8.2~MHz at this temperature.

\end{document}